\documentclass[aps, prd, amsmath, amssymb, amsfonts, floats, %
floatfix, superscriptaddress, 
groupaddress,nofootinbib,10pt,twocolumn]
{revtex4-1}

\usepackage[tracking=true]{microtype}
\SetTracking{}{500}
\SetTracking{encoding={*}, shape=sc}{40}

\usepackage{graphicx}
\usepackage{subfigure}
\usepackage{color}
\usepackage{soul}
\usepackage{booktabs}
\usepackage[USenglish]{babel}

\usepackage{graphics,color}
\usepackage{bm}
\usepackage{amsmath}
\usepackage{multirow}

\def\be{\begin{equation}}
\def\ee{\end{equation}}
\def\ba{\begin{eqnarray}}
\def\ea{\end{eqnarray}}

\usepackage[T1]{fontenc}
\usepackage{tikz}
\usetikzlibrary{positioning,decorations.pathmorphing}
\usepackage{pgf}
\usepackage{pgffor}
\usepgfmodule{shapes}
\usepgfmodule{plot}
\usetikzlibrary{decorations}
\usetikzlibrary{arrows}
\usetikzlibrary{snakes}

\tikzstyle{cnode}=[circle, draw, thin,fill=white!20, scale=0.7]

\newcommand{\bL}{\bar{\Lambda}}
\newcommand{\sL}{\mathcal{L}}
\newcommand{\sM}{\mathcal{M}}
\newcommand{\bJ}{\bar{\mathcal{J}}}

\begin{document} 
 
\title{Rotating black holes in three-dimensional Ho\v{r}ava gravity} 
\author{Thomas P. Sotiriou}
\affiliation{School of Mathematical Sciences and School of Physics and Astronomy,
University of Nottingham, University Park, Nottingham, NG7 2RD, United Kingdom}

\author{Ian Vega}
\affiliation{SISSA, Via Bonomea 265, 34136, Trieste, Italy and INFN, Sezione di Trieste, Italy}

\author{Daniele Vernieri}
\affiliation{SISSA, Via Bonomea 265, 34136, Trieste, Italy and INFN, Sezione di Trieste, Italy}

\begin{abstract}
We study black holes in the infrared sector of three-dimensional Ho\v{r}ava gravity. It is shown that black hole solutions with anti-de Sitter asymptotics are admissible only in the sector of the theory in which the scalar degree of freedom propagates infinitely fast. We derive the most general class of stationary, circularly symmetric, asymptotically anti-de Sitter black hole solutions. We also show that the theory admits black hole solutions with de Sitter and flat asymptotics, unlike three-dimensional general relativity. For all these cases, universal horizons may or may not exist depending on the choice of parameters. Solutions with de Sitter asymptotics can have universal horizons that lie beyond the de Sitter horizon.
\end{abstract} 
\maketitle  
 
\section{Introduction}

The celebrated Banados-Teitelboim-Zanelli (BTZ) solution \cite{Banados-etal-PRL:1992,Banados-etal-PRD:1993} is the unique black hole of general relativity in three dimensions. For a number of years, such a  black hole solution was deemed impossible, and for good reason. In three-dimensional general relativity, there are no local gravitational degrees of freedom: curvature is algebraically fixed by the matter content, which implies that in a true vacuum the spacetime can only be flat. With a nonvanishing cosmological constant,  the field equations admit locally de Sitter (dS) and anti-de Sitter (AdS) solutions, but still preclude solutions with nontrivial curvature. Hence, one is led to believe that black hole solutions in three-dimensional Einstein gravity are impossible. 

This argument is evaded by noting that it relies solely on local considerations.  Taking account of its global structure, a spacetime can contain a black hole in spite of being locally maximally symmetric. The BTZ solution is an example; it is locally AdS, but it is turned into a black hole spacetime by certain identifications of spacelike related events. Ever since its discovery, the BTZ black hole has generated a considerable amount of attention, in large part due to its foreseen applications, particularly in addressing conceptual issues of quantum gravity that become more tractable in three dimensions.

This unrelenting desire for a quantum theory of gravity has also inspired a number of interesting theoretical ideas. One of these is the suggestion that Lorentz invariance may not be fundamental or exact, but is merely an emergent symmetry on sufficiently large distances or low energies. It has been suggested in Ref.~\cite{Horava:2009uw} that giving up Lorentz invariance by introducing a preferred foliation and terms that contain higher-order spatial derivatives can lead to significantly improved UV behavior. The corresponding gravity theory is dubbed Ho\v rava gravity and we will review its basic properties in the next section.

One needs to rethink the notion of a black hole in the presence of Lorentz violations. Theories that do not respect local Lorentz symmetry may propagate superluminal excitations that can penetrate the usual horizon. They may even have instantaneous (infinitely fast) modes, as is the case for Ho\v rava gravity. However, studies of four-dimensional black holes in this theory \cite{Barausse-etal-PRD:2011,blas-sibiryakov-prd:2011} have revealed the existence of a new type of horizon, called the universal horizon, which act as a causal boundary for all modes, irrespective of how fast they propagate. This allows one to generalize and preserve the (or some) notion of a black hole in the framework of Lorentz-violating gravity theories (see Ref.~\cite{Barausse:2013nwa} for a recent review).

In this paper, we focus on the infrared limit of three-dimensional Ho\v{r}ava gravity \cite{Sotiriou:2011dr} and our main goal is to seek a Lorentz-violating version of the BTZ black hole, i.e.~a black hole solution with AdS asymptotics. We show that such a spacetime exists only if one tunes the parameters of the theory. We find the most general family of solutions for this sector, assuming stationarity and circular symmetry, so generically our solution represents a rotating black hole. Remarkably, some of our solutions represent black holes even for a positive or vanishing cosmological constant.

Our motivation for seeking black hole solutions in the three-dimensional version of Ho\v{r}ava gravity is twofold. First, we hope that they will be useful as a playground for studying quantum field theory and quantum gravity effects in black hole spacetimes, as has been the case for the BTZ black hole. Second, we hope that we will gain some insight into the causal structure of black holes in the presence of Lorentz violations --- at least the aspects that do not depend on the dimensionality. It is worth stressing that our solutions are explicit and exact, unlike their four-dimensional counterparts. Indeed, the static, spherically symmetric solutions of Ref.~\cite{Barausse-etal-PRD:2011} are numerical and those of Ref.~\cite{blas-sibiryakov-prd:2011} are numerical and valid in the small-coupling limit. Explicit solutions in 4 dimensions are also known for specific, tuned choices of the parameters of the theory, but they are all static \cite{Berglund:2012bu}. The only rotating solutions currently known in four dimensions are not entirely explicit, and moreover, rely on the assumption of slow rotation \cite{Barausse:2012ny,Barausse:2012qh}. Working in three dimensions allows us to avoid approximations or numerics.

Throughout this paper, we adopt a mostly-minus spacetime signature and set $c =1$. In what follows, we will be referring to any solution for which the metric that couples minimally to matter fields has a Killing horizon as a black hole. We choose to do so because, for the matter fields, which we assume to be relativistic, the Killing horizon will be an event horizon. Hence, the spacetime will be a black hole in the conventional (GR) sense. It should be clear, however, that this is actually an abuse of terminology in the context of Ho\v rava gravity, as perturbations that reside in the gravity sector can propagate infinitely fast, as mentioned earlier, so one could have chosen to reserve the term black hole for solutions that have a universal horizon. 

\section{Ho\v{r}ava gravity and Einstein-aether theory}

In its original formulation, Ho\v{r}ava gravity is expli\-citly noncovariant and written in a preferred foliation. Introducing the line element  
\begin{equation}
ds^2 = N^2 dT^2 - g_{ij}(dx^i + N^idT)(dx^j + N^jdT), 
\end{equation}
where $N$ is the lapse and the leaves of the foliation are constant-$T$ hypersurfaces with induced metric $g_{ij}$.
The action of the theory in three dimensions then has the form \cite{Sotiriou:2011dr}
\begin{equation}
S_{\rm H} = \frac{1}{16\pi G_{H}} \int dT d^2x \sqrt{g} N \left[L_2+L_4\right], 
\label{eqn:hlaction}
\end{equation} 
where $G_H$ is a coupling constant with dimensions of a length squared, 
\begin{equation}
L_2=K_{ij}K^{ij} -\lambda K^2 + \xi \left({}^{(2)}R-2\Lambda\right) + \eta a_ia^i\,,
\end{equation}
$g$ is the determinant of the induced metric $g_{ij}$ on the constant-$T$ hypersurfaces, ($K_{ij}, K,{}^{(2)}R$) are its extrinsic, mean and scalar curvatures, respectively, and $a_i = -\partial_i \ln N$. $L_4$ collectively denotes a set of all terms with 4 spatial derivatives that are invariant under diffeomorphisms that leave the foliation untouched. 

The presence of these higher-order terms is crucial for improved UV behavior. Power-counting renormalizability requires terms of order $2d$ to be present \cite{Horava:2009uw}, where $d$ is the number of spatial dimensions. The full list of such terms for $d=2$ is given in Ref.~\cite{Sotiriou:2011dr}. The version of Ho\v rava gravity we are considering here is the most general one, without any further symmetries or assumptions. It is the three-dimensional counterpart of the 4-dimension action presented in Ref.~\cite{Blas:2009qj}. 

For what follows we will focus on the infrared limit of the theory by neglecting the $L_4$ terms. This is expected to be a good approximation so long as the curvature remains small enough and the foliation is sufficiently smooth.   

The low-energy part of the theory can be formulated in a covariant fashion, and it then becomes equivalent to a restricted version of Einstein-aether theory  \cite{Jacobson:2010mx}, which is a theory that couples the metric to a timelike, unit-norm vector field, $u_\alpha$, called the aether. The correspondence with Ho\v rava gravity is realized by restricting the aether to be hypersurface-orthogonal, or more specifically, normal to the constant-$T$ hypersurfaces. Hence,
\begin{equation}
    u_\alpha = \frac{\partial_\alpha T}{\sqrt{g^{\mu\nu}\partial_\mu T\partial_\nu T}}.
    \label{eqn:khronon}
\end{equation}
In (2+1) dimensions, Einstein-aether theory with a cosmo\-logical constant $\Lambda$ is defined by the action
\begin{equation}
    S_{\scriptsize\mbox{\ae}} = \frac{1}{16\pi G_{\scriptsize\mbox{\ae}}}\int d^3x \sqrt{-g} \left(-R -2\Lambda+ L_{\scriptsize\mbox{\ae}}\right)\,,
    \label{eqn:action}
\end{equation}
where $G_{\scriptsize\mbox{\ae}}$ is a coupling constant with dimensions of a length squared, $g$ is the determinant of $g_{\mu\nu}$, $\Lambda$ is the cosmological constant, $R$ is the 3D Ricci scalar,
\begin{equation}
    L_{\scriptsize\mbox{\ae}} = -M^{\alpha\beta\mu\nu}\nabla_\alpha u_\mu \nabla_\beta u_\nu\,,
\end{equation}
and 
\begin{equation}
    M^{\alpha\beta\mu\nu} = c_1 g^{\alpha\beta}g^{\mu\nu} + c_2g^{\alpha\mu}g^{\beta\nu} + c_3g^{\alpha\nu}g^{\beta\mu} +
    c_4 u^\alpha u^\beta g^{\mu\nu}.
\end{equation}
By giving up part of the gauge freedom and choosing $T$ as the time coordinate, the aether takes the form $u_\mu=N \delta^t_\mu$ and the action (\ref{eqn:action}) reduces to that of Ho\v rava gravity in the infrared limit, with the correspondence of parameters
\begin{equation}
	\label{eqn:corresp}
\frac{G_H}{G_{\scriptsize\mbox{\ae}}}=\xi = \frac{1}{1-c_{13}}, \hspace{2em} \frac{\lambda}{\xi} = 1 + c_2, \hspace{2em} \frac{\eta}{\xi} = c_{14},
\end{equation}
where $c_{ij} = c_i+c_j$. 

In the covariant formulation of the theory the preferred time $T$ becomes a scalar field that defines the preferred foliation at the level of the solution. Irrespective of the formulation, the theory propagates a scalar degree of freedom in three dimensions and there is no spin-2 graviton \cite{Sotiriou:2011dr}. It is important to stress that when the $L_4$ terms are ignored, the scalar mode will have a linear dispersion relation in flat space, whereas, in the full theory the dispersion relation will be rational and well approximated by $\omega^2\sim k^4$ for large momenta. So, excitations with sufficiently high momenta can reach arbitrarily high speeds. Moreover, the theory has an instantaneous mode even in the low-energy limit  (see Ref.~\cite{blas-sibiryakov-prd:2011} for a discussion in four dimensions). Both of these facts are particularly relevant for black hole spacetimes. High-energy modes will be able to penetrate surfaces that appear as usual horizons in the low-energy limit of the theory. More importantly, even within the framework of the low-energy approximation, the presence of instantaneous, infinite speed, modes means that information can be transmitted through these horizons.

\section{Reduced action}

We find it convenient to work with the covariantized version of Ho\v rava gravity, equations of motioni.e.~Einstein-aether theory with the aether assumed to be hypersurface orthogonal {\em before} the variation. Assuming stationarity and circular symmetry, the most general metric in (2+1) dimensions is given by 
\begin{equation}
    ds^2 = Z(r)^2 dt^2 -\frac{1}{F(r)^2}dr^2 -r^2(d\phi+\Omega(r)dt)^2\,.
    \label{eqn:btzansatz}
\end{equation}
The aether field is also just a function of $r$: $u_\alpha(x^\beta)=u_\alpha(r)$. We shall refer to these as BTZ coordinates.

In three dimensions, $u_\alpha$ is hypersurface-orthogonal if and only if $u_{[\alpha}\nabla_\beta u_{\gamma]}=0$, which in BTZ coordinates is explicitly, $u_t \partial_r u_\phi = u_\phi \partial_r u_t$. A trivial solution to this is $u_\phi=0$. 
More generally, the hypersurface-orthogonality condition can be integrated to give $u_\phi = Cu_t$, for some constant $C$. This must hold throughout the spacetime. If $C\neq 0$, we see from Eq. (\ref{eqn:khronon}) that $T$ will satisfy $\partial_{\phi}T= C\partial_tT$. This means that the dependence of $T$ on $t$ and $\phi$ can be only through the combination $\zeta = t+C\phi$. In other words, we have $T(t,r,\phi) = f(r,\zeta)=f(r,t+C\phi)$, for some arbitrary function $f(r,\zeta)$. But the coordinate $\phi$ runs along orbits of the spacelike axial Killing vector of the spacetime. Keeping all other coordinates fixed, there must then exist a constant $p$ so that $\phi$ and $\phi+p$ refer to the same spacetime event. This means that $f(r,\zeta)$ will either be multivalued on each spacetime event, or it will have to be periodic in both $\phi$ and $t$. None of these options seem to be acceptable for a coordinate that is supposed to act as the preferred time of a global foliation. Hence, 
we shall only focus on aether configurations for which $C=0$ or $u_\phi=0$.

With $u_\phi=0$,  the unit norm constraint allows us to parametrize the aether as
\begin{eqnarray}
    u_t = \pm \sqrt{Z(r)^2(1+F^2(r)U^2(r))}, \hspace{1em} u_r = U(r),
    \label{eqn:aetheransatz}
\end{eqnarray}
where we denote $u_r$ by the function $U(r)$ from now on. With no loss of generality\footnote{Choosing the alternative (--) branch yields the same reduced action.}, we shall choose the positive (+) branch for $u_t$.

Inserting Eqs. (\ref{eqn:btzansatz}) and (\ref{eqn:aetheransatz}) into Eq. (\ref{eqn:action}), discar\-ding boundary terms, and using the Ho\v{r}ava parameters $\{\lambda, \xi,\eta\}$, we arrive at the reduced action
\begin{equation}
    S_{\rm r} = \frac{1}{8G_H} \int dt dr L_{\rm r},
    \label{eqn:reduced}
\end{equation}
where 
\begin{widetext}
\begin{eqnarray}
	L_{\rm r} =  &\frac{r^3
   F}{2 Z}({\Omega}')^2 - 2\xi Z \left(\Lambda \frac{r}{F} + F'\right)+\frac{r \eta  FZ'^2}{Z} 
   +\frac{(1-\lambda ) F^3 Z U^2}{r}
  +\frac{r F Z \Big(1-\lambda +(1+\eta -\lambda ) F^2 U^2\Big)}{1+F^2 U^2} \left(U F'+F
   U'\right)^2\nonumber \\ 
   & +r (1+\eta -\lambda ) F^2 U Z' \left(U \left(2 F'+\frac{F Z'}{Z}\right)+2 F
   U'\right) +2 (\xi-\lambda) F^2 U \Big(F U Z'+Z \left(U F'+F U'\right)\Big).
   \label{eqn:reducedL}
\end{eqnarray}
\end{widetext}
Requiring stationarity of the reduced action, $\delta S_{\rm r}=0$, then supplies our equations of motion. These are the Euler-Lagrange (EL) equations with respect to the functions $Z, F, \Omega$ and $U$. 

Results obtained with the reduced action approach should always be treated and interpreted with some caution. Critical points with respect to symmetric variations of the action need not be stationary points with respect to general variations. Therefore, solutions to equations of motion that arise from symmetry-reduced actions need not satisfy the full field equations \cite{Palais:1979,Fels:2001rv,Deser:2003up,Deser:2004yh,Deser:2004gi}. However, \emph{any} symmetric solution to the full field equation ought to be a critical point with respect to symmetric variations. The equations of motion from symmetric variations then constitute \emph{necessary} conditions for any solution to the full field equations.  If one succeeds in integrating them (or a subset of them), one can simply check if the solutions indeed satisfy the full field equations \cite{Deser:2003up}. This is the strategy we adopt here.

\section{$\Omega$ equation}

From Eq. (\ref{eqn:reducedL}), the EL equation with respect to $\Omega$ is
\begin{equation}
{\Omega}'' +\left(\frac{3}{r}+\frac{F'}{F}-rFZ'\right){\Omega}'= 0.
\end{equation}
This can be integrated to give 
\begin{equation}
    \Omega(r) = c+ \mathcal{J}\int^r \frac{Z(r')}{{r'}^3F(r')} dr',
    \label{eqn:nphi}
\end{equation}
for integration constants $\mathcal{J}$ and $c$. 

With the coordinate transformation $\{t\rightarrow t', \phi \rightarrow \phi'-ct'\}$, we can set  $c=0$ without loss of generality. 
Substituting $\Omega$ into each of three remaining EL equations, we are left with a coupled nonlinear system in the remaining unknowns $\{Z,F,U\}$, which are too lengthy to be usefully displayed here. In the remainder, we refer to the EL equation corresponding to $Z$ as the $Z$ equation, and likewise for the others.

\section{anti-de Sitter and asymptotically anti-de Sitter solutions}
\label{sec:maxSymm}

A natural starting point is to look for maximally symmetric solutions in three-dimensional Ho\v{r}ava gravity. After all, the (BTZ) black hole of three-dimensional general relativity belongs to this class of spacetimes (i.e., AdS), and we shall search for solutions that approach BTZ in the appropriate limit. We shall discover in this section that any asymptotically AdS analogue in  Ho\v rava gravity can only exist in the $\eta = 0$ sector of the theory (see also Ref.~\cite{Janiszewski:2014iaa}).

In three dimensions, a spacetime is (locally) maximally symmetric if
\begin{equation}
		M_{\mu\nu}:=R_{\mu\nu} -\frac{R}{3}g_{\mu\nu} = 0.
		\label{eqn:maxsymm}
\end{equation} 
When $F$ does not vanish identically, one finds that $M_{rr}=0$ and $M_{\phi\phi}=0$ can be combined to give
\begin{equation}
	Z(FZ'-ZF')=0,
\end{equation}
from which we conclude that $Z=\kappa F$ is necessary for maxi\-mal symmetry. $\kappa$ is some constant, which we can always set to 1 without loss of generality by a time rescaling.

Now, inserting $Z=F$ into Eq. (\ref{eqn:nphi}), which is one of the EL equations, we get
\begin{equation}
    \Omega = -\frac{\mathcal{J}}{2r^2}. 
\end{equation}
This in turn reduces all of $M_{\alpha\beta}=0$ into  a single differential equation, 
which can be integrated to give
\begin{equation}
	F^2 = Z^2 = \frac{\mathcal{J}^2}{4r^2}+A+Br^2. \label{eqn:maxsymm} 
\end{equation}
For such metrics, the scalar curvature is $-6B$. The geometry is either dS, AdS or flat\footnote{Note, however, that when $A\neq 1$, these spacetimes have a deficit angle. The literature has referred to these as ``quasi-asymptotically flat,'' but for convenience, we shall call them simply ``flat.''}, when $B < 0$, $B > 0$, or $B=0$, respectively.

With Eq. (\ref{eqn:nphi}) being a necessary condition, Eq. (\ref{eqn:maxsymm}) can be taken to be the most general form of a maximally symmetric spacetime in three-dimensional Ho\v{r}ava gravity. In what follows, we shall discover black hole solutions very similar in form. 

To check whether metrics of this form indeed exist in three-dimensional Ho\v{r}ava gravity and, if so, to specify their corresponding aether configurations, we return to the EL equations. Since $Z=F$, these now form a coupled system of three nonlinear differential equations for $\{F(r),U(r)\}$. For a solution to exist, these equations clearly must not all be independent of each other. 

By systematically eliminating terms proportional to $\xi$ and $\lambda$ in the EL equations, they can be combined to give the equation 
\begin{equation}
\eta u_t^3(u_t'+ru_t'') = 0.
\label{eqn:utconstraint}
\end{equation}
If $\eta \neq 0$, then we have $u_t(r) = c + d\ln r$, where $c$ and $d$ are arbitrary integration constants. Using Eq. (\ref{eqn:aetheransatz}), one gets
\begin{equation}
U(r) = \pm \sqrt{\frac{(c+d\ln r)^2-F(r)^2}{F(r)^4}}.
\end{equation}

For AdS space, which is our primary interest here, we have $F^2 \sim \alpha r^2$ with $\alpha > 0$, which clearly leads to an ill-defined aether because $r^2 \gg \ln r$ as $r \rightarrow \infty$. 
We conclude from all this that AdS is a not a solution in three-dimensional Ho\v{r}ava gravity when $\eta \neq 0$. 

The restriction $Z=F$ might seem overly restrictive if we only want to require that the spacetime be AdS only asymptotically. In BTZ coordinates, boundary conditions for asymptotically AdS spacetimes in three dimensions were previously identified in \cite{brown-henneaux:1986}. These read 
\begin{equation}
\begin{aligned}
g_{tt} &= \frac{r^2}{\mathcal{L}^2} + O(1) \\
g_{tr} &=  O(r^{-3})  \\
g_{t\phi} &= O(1) \\
g_{rr} &= -\frac{\mathcal{L}^2}{r^2} + O(r^{-4})  \\
g_{r\phi} &= O(r^{-3}) \\
g_{\phi\phi} &= r^2 +O(1), 
\end{aligned}
\end{equation}
where $\mathcal{L}$ is the length scale associated with the asymptotic curvature, which is specified by an effective cosmological constant, $\bL = -1/\mathcal{L}^2$.
These require that our metric functions behave asymptotically as
\begin{equation}
\begin{aligned}
\Omega &= O(r^{-2}) \\
Z &=  \frac{r}{\mathcal{L}}+O(r^{-1})  \\
F &= \frac{r}{\mathcal{L}} + O(r^{-1}). 
\end{aligned}
\end{equation}
The solution for $\Omega$ in Eq.  (\ref{eqn:nphi}) satisfies this. Now if $U \sim U_0r^m$ as $r\rightarrow \infty$, for some unspecified $m$, then the leading-order terms in the EL equations cannot simultaneously vanish unless $m = -1$ or $m=-3$. More importantly, for either choice of fall-off, it can be shown that $\eta$ has to be zero. A lengthy but straightforward demonstration can be found in Appendix \ref{app:asympAdS}. We show in Appendix \ref{app:align} that when $m=-1$, the aether is not orthogonal to constant-$t$ surfaces (i.e. it does not align with the timelike Killing vector) asymptotically, but this does happen when $m=-3$.

\section{Black hole solution for $\eta = 0$}

\subsection{The solution}

The considerations of the previous section suggest that, in looking for a BTZ analogue, we ought to focus on the $\eta=0$ sector of the theory. In this sector, the EL equations take the generic form 
\begin{align}
    (\lambda-1) (FUZ'' &+  FZU'' + ZUF'') \nonumber \\  &+ \mathcal{H}(Z,Z',F,F',U,U') = 0, 
\end{align}
with $\mathcal{H}$ being a nonlinear algebraic function of the unknowns and their derivatives. One way to simplify the problem would be to choose $U=0$, which would mean choosing a configuration in which aether is globally aligned with the timelike Killing vector. This approach was followed in Ref.~\cite{Park:2012ev} and parts of Ref.~\cite{Shu:2014eza}\footnote{Ref.~\cite{Shu:2014eza}, which appeared during the late stages of the preparation of this manuscript, contains a collection of static (nonrotating) solutions for special values of the parameters $\xi$, $\lambda$, and $\eta$ and/or restrictions in the metric ansatz. These special choices seem to be motivated by the fact that they simplify the calculations and make it easier to obtain analytic solutions. The diagonal solutions in the preferred foliation are not black holes for the reasons discussed above. The causal structure of the nondiagonal solutions and the behavior of the corresponding foliation is left unexplored in Ref.~\cite{Shu:2014eza}.} by working directly in the preferred foliation. (We discuss the correspondence of the two approaches in Appendix \ref{app:aligned}). Impo\-sing global alignment trivializes the $U$ equation and kills all second-order derivatives in the remaining EL equations, paving an easier route to exact solutions. However, it is easy to argue that these solutions cannot represent black holes in Ho\v{r}ava gravity. The Killing vector ($\partial_t$) is null at the Killing horizon (or the ergosurface) and spacelike inside it, but the aether has to be timelike everywhere if it is to define a foliation by spacelike hypersurfaces of constant preferred time. Global alignment is thus kinematically impossible in black hole spacetimes.

Without any a priori assumptions about $U$, the EL equations can nevertheless be combined to give
\begin{equation}
\frac{4\xi}{\Lambda} r^3 Z F \left(Z F'-FZ'\right) = 0.
\end{equation}
Since we wish to keep other coupling constants generic, and since neither $Z$ nor $F$ vanish identically, we can conclude that $\eta=0$ necessitates $Z=\kappa F$, where again we shall set $\kappa =1$ with no loss of generality. 

Using this, the $U$ equation turns into
\begin{align}
   (\lambda-1) &\bigg\{ (FU''+2UF'') + \frac{1}{r^2F} \big[2 r^2 U {F'}^2 \nonumber\\ &+2 r F F' (2 r U'+U) +F^2 (r
    U'-U)\big]\bigg\} = 0,
    \label{eqn:ueqn}
\end{align}
and the $Z$ and $F$ equations collapse into a single equation (which we shall not display here due to its length). 

With the change of variables,
\begin{equation}
    y = UF^2, 
    \label{eqn:subs}
\end{equation}
Eq. (\ref{eqn:ueqn}) turns into the simple differential equation
\begin{equation}
    r^2\frac{d^2y}{dr^2} + r\frac{dy}{dr} -y=0.\label{eqn:simple}
\end{equation}
More geometrically, this equation means that constant-$T$ surfaces have constant mean curvature.\footnote{The fact that constant preferred time surfaces have constant mean curvature is also a property of Cuscuton theory, which has been argued to be related to Ho\v{r}ava gravity \cite{Afshordi:2006ad,Afshordi:2009tt}.} (See Appendix \ref{app:meancurv}). The general solution to Eq. (\ref{eqn:simple}) is 
\begin{equation}
    y= UF^2 = \frac{a}{r} + br,
\end{equation}
where $a$ and $b$ are integration constants. 
Therefore, $U$ and $F$ have to be related in the following way:
\begin{equation}
    U = \frac{1}{F^2} \left(\frac{a}{r} + br\right).
    \label{eqn:fol}
\end{equation}
Inserting this into either the $Z$ or $F$ equation, we get
\begin{align}
    \frac{1}{2}\frac{d}{dr}\left(F^2\right) &+
    \left[\frac{\mathcal{J}^2+4a^2(1-\xi)}{4\xi}\right]\frac{1}{r^3} \nonumber \\ &-
    \left[b^2\left(\frac{2\lambda-\xi-1}{\xi}\right)-\Lambda\right]r = 0.
\end{align}
This leads to the metric functions
\begin{subequations}
\label{eqn:soletazero}
\begin{align}
    F^2 = Z^2 &= -\mathcal{M} +\frac{\bar{\mathcal{J}}^2}{4 r^2}-\bar{\Lambda}r^2 \label{eqn:soletazero1}\\
       \Omega &= -\frac{\mathcal{J}}{2r^2},
       \label{eqn:soletazero2}
       \end{align}
\end{subequations}
where $\sM$ is an integration constant and 
\begin{align}
    \bar{\mathcal{J}}^2 &= \frac{\mathcal{J}^2+4a^2(1-\xi)}{\xi}\\
          \bar{\Lambda} &= \Lambda - \frac{b^2(2\lambda-\xi-1)}{\xi}. \label{eqn:effLambda}
\end{align}
In the limit to general relativity ($\lambda \rightarrow 1, \xi \rightarrow 1$), Eq. (\ref{eqn:soletazero}) gives the familiar BTZ metric. When $\xi=1$, and thus $\bar{\mathcal{J}}=\mathcal{J}$, the solution becomes the BTZ metric with a shifted cosmological constant, $\bar{\Lambda}=\Lambda -2b^2(\lambda-1)$. Note that $\bar{\mathcal{J}}^2$ \emph{can} be negative; this happens when either $\xi <0$ or $\xi > 1, a^2 > \mathcal{J}^2/(4(\xi-1))$. 

The aether configuration for this metric is
\begin{subequations}
\label{eqn:aetheretazero}
\begin{eqnarray}
u_r &=& \frac{1}{F^2}\left(\frac{a}{r}+br\right), \label{eqn:aetheretazeroR}\\
u_t &=& \sqrt{F^2+\Big(\frac{a}{r}+br\Big)^2}.
\label{eqn:aetheretazeroT}
\end{eqnarray}
\end{subequations}
Since a vanishing $u_r$ signifies alignment of the aether with the timelike Killing vector, the constants $a$ and $b$ can be regarded as measures of aether misalignment. Of these two aether parameters, $b$ is what dominates asymptotically and is what affects the asymptotic behavior of the metric. As shown in Appendix {\ref{app:align}}, if $b \neq 0$, then the aether does not  align with the timelike Killing vector asymptotically. Thus, the parameter $b$ can be understood to be a measure of asymptotic misalignment.

Taken together, Eqs.~(\ref{eqn:soletazero}) and (\ref{eqn:aetheretazero}) give the most general metric and aether configuration in the $\eta = 0$ sector. It is a four-parameter family of solutions, specified by $\{\mathcal{M},\mathcal{J}, a, b\}$. 

Unless one imposes restrictions on the parameters, $u_t$ can become imaginary in parts of the spacetime. That would signal a breakdown of the foliation. It is reasonable to restrict one's attention to solutions for which a foliation exists all the way to the singularity, since the existence of a well-defined spacelike foliation is essential in Ho\v rava gravity. This can be achieved by imposing the condition $F^2 +(a/r +br)^2 >0$ or
\begin{align}
\frac{1}{r^2}\Big((b^2 -\bL)r^4 + (2ab-\sM)r^2 + \left(\frac{\bJ^2}{4}+a^2\right) \Big)> 0\,.
\label{eqn:realaether}
\end{align} 
As $r\rightarrow 0$, the combination $a^2 + \bJ^2/4$ or $(\mathcal{J}^2/4+a^2)/\xi$ dominates $u_t^2$, and so it must be positive. Thus, in order to ensure the existence of a foliation close to the singularity, we are restricted to working in the domain $\xi>0$.

At large $r$, the term whose coefficient is $(b^2-\bL)$ dominates instead. This coefficient is always positive for AdS asymptotics, as $\bL <0$. For dS asymptotics one would have to impose that $b^2 > \bL$ in order for the foliation to not end at some finite $r$.

\subsection{Curvature scalars and asymptotics}

A quick calculation of the scalar curvature\footnote{Note that because of our convention, AdS space ($\bar{\Lambda}<0$) gives $R>0$.} gives
\begin{equation}
	R = -6\bar{\Lambda} + \frac{1}{2r^4}\left(\bar{\mathcal{J}}^2-\mathcal{J}^2\right),
\end{equation}
which is not constant and generically diverges at $r=0$. When $\xi = 1$, we have $\bar{\mathcal{J}} = \mathcal{J}$, so the Ricci scalar is constant, but it can be of either sign depending on $\lambda, \Lambda$, and $b$. The Kretschmann scalar also diverges at $r=0$:
\begin{align}
R_{\alpha\beta\gamma\delta}R^{\alpha\beta\gamma\delta} = 12 \bar{\Lambda}^2 &-\frac{2\bar{\Lambda}}{r^4}\left(\bar{\mathcal{J}}^2-\mathcal{J}^2\right) \nonumber \\ 
&+\frac{11}{4r^8}\left(\bar{\mathcal{J}}^2-\mathcal{J}^2\right)^2.
\end{align}
These imply that $r=0$ is a curvature singularity, unless $\bar{J}^2=J^2$. This is in contrast to the BTZ black hole for which $r=0$ is neither a curvature nor a conical singularity, but is instead a ``causal'' singularity where both the Ricci and Kretschmann scalars are finite and perfectly smooth. 

The metric can be (quasi) asymptotically flat, dS, or AdS, irrespective of the sign of the (bare) cosmological constant, $\Lambda$ (which will be negative, $\Lambda = -1/l^2$, for BTZ). The sign of the effective cosmological constant,
\begin{equation}
\bar{\Lambda} = \Lambda-\frac{b^2(2\lambda-\xi-1)}{\xi},
\label{eqn:effLambda2}
\end{equation}
determines the asymptotic behavior of the metric. 

\subsection{Setting $\xi=1$ by redefinitions}

\label{xi=1}

It is clear that $\xi=1$ is a special value for the solution we have found. The metric reduces to the BTZ solution and the curvature singularity disappears.

However, one can actually set $\xi=1$ by means of field redefinitions. In the preferred frame picture, one can perform a constant rescaling of the lapse function $N$. If one sets the new lapse $N'=\sigma N$, action (\ref{eqn:hlaction}) (always restricting attention on the $L_2$ part only) remains invariant apart from an overall factor and after the following parameter rescaling
\begin{align}
\xi' &= \frac{\xi}{\sigma}\\
\eta' &= \frac{\eta}{\sigma} \\
\lambda' &= \lambda\\
\Lambda'&=\frac{\Lambda}{\sigma}\,.
\end{align}
This implies that, with the choice $\sigma=\xi$, any theory in the sector $\{\eta=0, \xi>0\}$ can be mapped onto $\{\eta=0,\xi=1\}$.

In the covariant picture, the corresponding redefinition is
\begin{align}
	g'_{\alpha\beta} &= g_{\alpha\beta}+(\sigma-1)u_\alpha u_\beta \\
	u'{}^\alpha &=\frac{1}{\sqrt{\sigma}}u^\alpha,
\end{align}
with the same rescaling for $\Lambda$ (where $\sigma$ is restricted to be positive so that the new metric is Lorentzian). This redefinition was first considered in Ref.~\cite{Foster:2005ec}. The action for the primed fields takes the same form as the action for the unprimed ones up to the values of the coefficients $c_i$. The primed action takes on coefficients $c'_i$ that are related to the $c_i$. These relations are such that
\begin{align}
	1+c'_2 &= \sigma(1+c_2) \\
	c'_{14} &= c_{14} \\
	c'_{13} - 1 &= \sigma(c_{13}-1).
\end{align}
Using the correspondence in Eq. (\ref{eqn:corresp}) one can verify that choosing $\sigma=\xi$ one can set $\xi$ to 1. 

Clearly, using these redefinitions allows one to work with a more familiar spacetime, which is free of curvature singularities. (It does not actually simplify the derivation of the solution significantly). However, we will choose not to follow this route. Such a redefinition is only allowed in vacuo. If other fields couple to the lapse, the shift and 3-metric (or the metric and the aether), then such a redefinition no longer leaves the action invariant. Additionally, one might be interested specifically in the spacetime structure of $g_{\mu\nu}$. For instance, in four dimensions one can require that $g_{\mu\nu}$ couples minimally to the matter in order for the equivalence principle to be satisfied. This would make this metric distinct. Here we are considering three dimensions, but if we want to use our solutions to understand something about three-dimensional black holes it seems prudent to understand the structure of $g_{\mu\nu}$ itself. As we will see later on, the causal structure of the two metrics can also be different.

\subsection{Metric horizons and causal structure}

\begin{table*}[!t]
\centering
\squeezetable
\renewcommand\arraystretch{1.3}
\begin{tabular}{@{}p{3cm}p{3cm}p{2.8cm}p{2.8cm}p{2.8cm}p{2.8cm}@{}}
\toprule
Case & {} & $\bJ^2\bL \geq 0$ & $\bJ^2 = 0$ & $0 > \bJ^2\bL > -\sM^2$ & $\bJ^2\bL= -\sM^2$ \\
\toprule
\multirow{2}{*}{$\sM >0,\,\, \bL <0$} & horizons & $\tilde{r}_+$ {\bf (b)} & $\sqrt{2}r_{(1/2)}$ {\bf (b)} & $r_\pm$ {\bf (b)} & $r_{(1/2)}$ {\bf (b)}\\
& singularity & spacelike & spacelike & timelike & timelike \\
\hline
\multirow{2}{*}{$\sM >0,\,\, \bL >0$} & horizons & $\tilde{r}_-$ {\bf (c)} &  {---} & {---} & {---} \\
& singularity & timelike &  spacelike & spacelike & spacelike \\
\hline
\multirow{2}{*}{$\sM <0,\,\, \bL <0$} & horizons & $\tilde{r}_-$ {\bf (b)} & {\bf ---} & {\bf ---} & {\bf ---} \\
& singularity & spacelike & timelike & timelike & timelike \\
\hline
\multirow{2}{*}{$\sM <0,\,\, \bL >0$} & horizons & $\tilde{r}_+$ {\bf (c)} & $\sqrt{2}r_{(1/2)}$ {\bf (c)} & $r_+$ {\bf (c)}, $r_-$ {\bf (b)} & $r_{(1/2)}$ {\bf (c)} \\
& singularity & timelike & timelike & spacelike & spacelike \\
\hline
\bottomrule
\end{tabular}
\renewcommand\arraystretch{1}
\caption{Killing horizons and the nature of the curvature singularity for various cases. Each of the Killing horizons is denoted either by a {\bf (c)} for de Sitter (cosmological) horizon, or {\bf (b)} for black hole (event) horizon. Their locations are specified by: $r_\pm^2 := |\sM/(2\bL)|\left(1\pm(1-|\bJ^2\bL|/\sM^2)^{1/2}\right), \tilde{r}_\pm^2 := |\sM/(2\bL)|\left((1+|\bJ^2\bL|/\sM^2)^{1/2}\pm1\right), r_{(1/2)}^2 := |\sM/(2\bL)|.$  $\bL =0$ is excluded from this table, simply because $\bJ^2\bL$ vanishes and the sign of $\bJ^2$ cannot be immediately inferred. In this case, an asymptotically flat black hole exists for $\{\sM =-1, \bJ^2 < 0\}$ and the horizon radius is $r_o=\sqrt{-\bJ^2}/2$.}
\label{table:horcaus}
\end{table*}

In stationary spacetimes, horizons are null, stationary surfaces. The normal to any stationary surface must be proportional to $\partial_\alpha r$, and this is null when 
$g^{\alpha\beta}(\partial_\alpha r)(\partial_\beta r) = g^{rr}=-F^2=0$, or \begin{equation}
g^{rr}=\frac{\bL}{r^2}(r^2-r_+^2)(r^2-r_-^2) =0 \,.
\end{equation} 
The horizons are thus located at $r=r_\pm$, where
\begin{equation}
r^2_\pm = -\frac{\mathcal{M}}{2\bar{\Lambda}}\left[1\pm \sqrt{1+\dfrac{\bJ^2 \bL}{\sM^2}}\right]. 
\label{eqn:existhor}
\end{equation}
For there to be two horizons (i.e. for both values in Eq. (\ref{eqn:existhor}) to be real), both $\sM/\bL$ and $\bJ^2\bL$ must at least be negative. In which case, we can write
\begin{equation}
\sM = -\bar{\Lambda}(r_+^2+r_-^2), \hspace{1cm} \bJ^2 = -4\bL (r_+r_-)^2.
\end{equation}
The case $\{\bL < 0, \sM >0, \bJ^2 >0\}$ corresponds closely to the BTZ solution of general relativity. For this BTZ-like branch of our solutions, there exists an analogous ``angular momentum" bound
\begin{equation}
\bar{\mathcal{J}}^2 \leq \frac{\mathcal{M}^2}{|\bar{\Lambda}|},
\label{eqn:bound}
\end{equation}
which guarantees that $r_\pm$ are both real. These are the locations of the inner and outer horizons of the black hole. When the bound is saturated, the horizons coincide at $r = r_{(1/2)} := \sqrt{\left|\sM/(2\bL)\right|}$. The inner horizon approaches $r=0$ when $\bJ^2 \rightarrow 0^+$, while keeping a fixed $\bL <0$. 
As $\bL \rightarrow 0^-$, while keeping $\bJ^2 >0$, $r_+$ gets pushed to infinity so that only the interior of the black hole remains. This is similar to the situation in three-dimensional general relativity, where the black hole can only be asymptotically AdS, because the relevant parameter is a strictly non-negative $\mathcal{J}^2$, rather than $\bJ^2$.

Remarkably, there exist solutions with black hole horizons and de Sitter or flat asymptotics.  In particular, when $\{\bL > 0, \sM < 0, \bJ^2 < 0\}$, $r_\pm$ are both still real and their associated hypersurfaces are both Killing horizons. But since $\bL >0$, $r_+$ corresponds to the dS horizon, and $r_-$ takes the role of the black hole event horizon. For $\{\bL =0, \sM =-1, \bJ^2 < 0\}$, one obtains an asymptotically flat black hole with a horizon at $r= r_o :=\sqrt{-\bJ^2}/2$ (for $\sM\neq -1$ the asymptotics would be ``quasi asymptotically flat''). In general relativity such solutions do not exist because $\mathcal{J}^2$ plays the role of $\bJ^2$, and $\mathcal{J}^2$ is strictly non-negative.

Other possibilities exist for which there is only one Killing horizon, which can be either an event horizon or a dS horizon, depending on the sign of the cosmological constant. Many of these cases are summarized in Table \ref{table:horcaus}, which also provides the respective positions of the Killing horizons for convenience. 

We note as well that these spacetimes can have ergoregions. These are demarcated by $r=r_{\rm erg}$, such that $g_{tt}(r_{\rm erg})= Z^2-r_{\rm erg}^2(\Omega)^2=0$. The ergosurfaces are thus located at
\begin{equation}
(r_{\pm}^{\rm erg})^2= -\frac{\mathcal{M}}{2\bar{\Lambda}}\left(1\pm \sqrt{1+\dfrac{\bL}{\sM^2}\Delta}\right).
\end{equation}
This is essentially Eq. (\ref{eqn:existhor}), with the replacement $\bJ^2 \rightarrow \Delta := \bJ^2 -\mathcal{J}^2$. The key parameter is then
\begin{equation}
\Delta:=\bJ^2 -\mathcal{J}^2 = \left(\frac{1-\xi}{\xi}\right)(\mathcal{J}^2+4a^2).
\end{equation}
When $\xi=1$, the ergosurface is uniquely at $r_{\rm erg} = \sqrt{\mathcal{M}/|\bar{\Lambda}|}$. We thus recover the BTZ case in general relativity for which $r_-\leq r_+\leq r_{\rm erg}$. In the parameter region $0< \xi \leq 1$ and for the BTZ-like case $\{\bL < 0, \sM >0, \bJ^2>0\}$, we have $\bJ^2> \Delta \geq 0$, and 
\begin{equation}
r^{\rm erg}_- \leq r_- \leq r_+ \leq r^{\rm erg}_+.
\end{equation} 
Outside the parameter region $0< \xi \leq 1$, $\Delta$ is negative, and $r^{\rm erg}_-$ becomes imaginary and so there is no ``inner'' ergosurface. Various other cases can be easily worked out, but they shall not be our concern for the rest of the paper.

Our next goal shall be to get a better sense of the spacetime's causal structure, for which we shall also need to know the character of its singularity, in addition to identifying its horizons and the nature of its asymptotic infinities. This is generally controlled by $\bJ^2$, whose sign dictates the behavior of $F^2$ as $r\rightarrow 0$.

Consider first the case $\bJ^2 \neq0$. Then as $r\rightarrow 0$, $g^{rr}= g^{\mu\nu}(\partial_\mu r)(\partial_\nu r) = -F^2 \sim -\bJ^2/(4r^2)$. The normal to constant-$r$ surfaces is then spacelike when $\bJ^2>0$ (like the rotating BTZ black hole) or timelike when $\bJ^2<0$. When $\bJ^2 =0$ and $\mathcal{J}^2 \neq 0$, there will still exist a curvature singularity, but whether it is timelike or spacelike now depends on the sign of $\mathcal{M}$, since  $g^{rr} \rightarrow \mathcal{M}$ as $r\rightarrow 0$. When $\mathcal{M}>0 \,\,(\mathcal{M}<0)$, the singularity is spacelike (timelike). The spacelike nature of $r=0$ in the positive-$\sM$ case corresponds to the nonrotating BTZ black hole.

We have already mentioned in the previous section that the causal structure of the redefined metrics that lead to $\xi=1$ is different from that of $g_{\mu\nu}$. This should be clear now, as, in a suitable coordinate system, the redefined metric takes the same form as $g_{\mu\nu}$ but with $\xi=1$, so it is always a BTZ spacetime (potentially with different asymptotics than those of $g_{\mu\nu}$). Consider, for example, the asymptotically flat black holes that were discussed above and assume $J=0$ and $b=0$ (to avoid divergence of the aether asymptotically). The redefinition will lead to flat spacetime with a nontrivial aether. 

\subsection{Foliation and universal horizons}

\subsubsection{Regularity of the aether}
In the previous section, we focused mainly on the geometry of our solution, that is, on the metric and its properties. The causal structure of this metric is what is relevant to matter degrees of freedom minimally coupled to it. The second half of the solution is the aether field, or more precisely, the foliation it specifies. 

We shall first look at how the aether behaves along the horizons in the maximal extension.  For this it is sufficient to follow the Kruskal construction that brings the line-element to the form
\begin{equation}
	ds^2 = \bar{\Omega}(r)^2 dU dV - r^2 (d\bar{\phi}^2+ N^\phi(r)dt^2), 
\end{equation}
in terms of null coordinates $U$ and $V$, where $t = t(U,V)$, $r=r(U,V)$, $\bar{\phi} = \bar{\phi}(\phi,t(U,V))$. 

Several charts are generally needed to cover the full manifold, depending on how many Killing horizons the spacetime has. Only one chart is needed for the asymptotically flat case $(\sM = -1, \bL = 0, \bJ^2 < 0)$, which in BTZ coordinates has 
\begin{equation}
	F^2 = \frac{1}{r^2}\left(r^2 - R_0^2\right),
\end{equation}
where $R_0 = \sqrt{|\bJ^2|}/2$. The standard Kruskal coordinates are then
\begin{subequations}
	\label{eqn:KruskalFlat}
\begin{align}
	U &= \mp e^{-\kappa u} \\
	V &= e^{\kappa v},
\end{align}
\end{subequations}
where $u = t-r^*$, $v = t+r^*$, $r^* := \int F^{-2} dr$, and $\kappa = 1/R_0 = 2/\sqrt{|\bJ^2|}$ is the surface gravity of the horizon. The upper sign (-) is for the region $r > R_0$ and the lower sign (+) is for $r < R_0$. In this case, $r$ depends on $U$ and $V$ through
\begin{equation}
	e^{2\kappa r}\left(\frac{\kappa r-1}{\kappa r +1}\right) = -UV,
\end{equation}
and $\bar{\Omega}(r) = (1/\kappa)(1+1/(\kappa r)) \exp(-\kappa r)$. These coordinates are clearly regular through the Killing horizon.

The aether has components 
\begin{equation}
	u_{U} = \frac{1}{2\kappa U} \left[\left(\frac{a}{r} + br\right) \mp \sqrt{ \left(\frac{a}{r} + br\right)^2 + F^2}\right],
\end{equation}
and 
\begin{equation}
	u_{V} = \frac{1}{2\kappa V} \left[\left(\frac{a}{r} + br\right) \pm \sqrt{ \left(\frac{a}{r} + br\right)^2 + F^2}\right],
\end{equation}
where the upper signs hold for the future-pointing solution, $u_t>0$, which we have chosen to work with in the text, while the lower signs hold for the past-pointing solution, $u_t < 0$, which we have hitherto disregarded. 

Close to $R_0$, one can verify that $F^2 \simeq 2\kappa (r-R_0)$ and $r^* \simeq (2\kappa)^{-1}\ln |\kappa(r-R_0)|$, which imply $F^2 \simeq -2UV$. Therefore, as $U \rightarrow 0, V \rightarrow 0$ we have
 \begin{equation}
 	u_{U} \simeq \frac{1}{2\kappa U} \left[h_0 \mp |h_0| \left(1 - \frac{UV}{h_0^2}\right)\right],
	\label{eqn:uUkruskal}
 \end{equation}
 and 
 \begin{equation}
 	u_{V} \simeq \frac{1}{2\kappa V} \left[h_0 \pm |h_0| \left(1 - \frac{UV}{h_0^2}\right)\right],
	\label{eqn:uVkruskal}
 \end{equation}
where $h_0 := a/R_0 + bR_0$, which we assume not to vanish. Moreover, we shall assume for now that $h_0 >0$.

For the future-pointing solution, we therefore have 
\begin{align}
	u_{U} &\simeq \frac{V}{2\kappa h_0}  \label{eqn:nearHorizonPlusA}\\
	u_{V} &\simeq \frac{h_0}{\kappa V}, \label{eqn:nearHorizonPlusB}
\end{align}
as $r \rightarrow R_0$. The future-pointing solution is thus regular at the future event horizon ($U=0$), but is divergent at the past event horizon ($V = 0$). This divergence arises because the foliation turns null. In the various Penrose diagrams, we mark the singularity of the aether with dashed lines.

On the other hand, the past-pointing solution behaves like
\begin{align}
	u_{U} &\simeq \frac{h_0}{\kappa U} \\
	u_{V} &\simeq \frac{U}{2\kappa h_0}, 
\end{align}
and is thus regular at the past event horizon ($V = 0$) but divergent at the future event horizon ($U = 0$). 

This analysis also applies to the AdS case. For this, at least two charts are needed, each respectively in the neighborhoods of the two Killing horizons. The Kruskal coordinates for the flat space case carry over exactly to the region containing the outer horizon, except that the surface gravity is now $\kappa_+ := -\bL(r_+^2-r_-^2)/r_+$. Clearly then, the future-pointing aether field is again regular at the future event horizon and singular at the past event horizon. Around the inner horizon, one installs the usual coordinates, $U_- = \mp \exp(\kappa_- u), V_- = - \exp(-\kappa_- v)$, where $\kappa_- := -\bL(r_+^2-r_-^2)/r_-$. In these coordinates, the future-pointing aether can be seen to diverge at $V_- = 0$ and to remain regular at $U_- = 0$. A pattern thus emerges where the future-pointing aether diverges along past event horizons ($V^*=0$) and is regular along future event horizons ($U^*=0$), where $\{U^*$,$V^*\}$ are the outgoing/ingoing Kruskal coordinates adapted to an arbitrary Killing horizon. This remark holds for dS spacetimes as well.

\subsubsection{Universal horizons}

In a gravitational theory with nonlinear dispersion relations, the event horizon relinquishes its role as an absolute causal boundary. In spherically symmetric spacetimes, this role is taken over by the \emph{universal horizon} \cite{blas-sibiryakov-prd:2011,Barausse-etal-PRD:2011}, which arises when a constant preferred time (constant-$T$) surface coincides with a constant-$r$ surface. This constant-$r$ surface will then act as a causal boundary because any sort of physical process is presumed to proceed in the direction of increasing $T$. Therefore, any constant-$r$ hypersurface that happens to coincide with a constant-$T$ surface (i.e., a leaf of the foliation) can only be crossed in one direction. 

Because $u_\phi=0$, there will be a universal horizon when 
\begin{equation}
\partial_\alpha r \propto u_\alpha,
\end{equation}
or equivalently, when $u_\alpha t^\alpha =0$, where $t^\alpha$ is the timelike Killing vector. For the class of solutions given by Eq.~(\ref{eqn:soletazero}), the universal horizon is given by the surface $r(x^\alpha)=r_u$, where $r_u$ satisfies
\begin{equation}
u_t^2 = F(r_u)^2 + \left(\frac{a}{r_u}+br_u\right)^2=0 
\label{eqn:univhorA}
\end{equation}
or
\begin{equation}
(b^2-\bar{\Lambda})r^4+(2ab-\mathcal{M})r^2+\left(a^2 + \frac{\bar{\mathcal{J}^2}}{4}\right) = 0.
\label{eqn:univhor}
\end{equation}
The roots are 
\begin{align}
(r_u^\pm)^2 &= \frac{\mathcal{M}-2ab}{2(b^2-\bar{\Lambda})} \pm \frac{1}{2(b^2-\bar{\Lambda})}\times\nonumber\\ &\bigg((\mathcal{M}-2ab)^2-(4a^2+\bar{\mathcal{J}}^2)(b^2-\bar{\Lambda})\bigg)^{1/2}. 
\label{eqn:rootUH}
\end{align}
If the discriminant is negative then the roots will be imaginary and there will not be any universal horizon. If the discriminant is positive  both roots in Eq. (\ref{eqn:rootUH}) will be real and distinct. But then there will exist a region, $r^-_u <r<r^+_u,$ where the aether turns imaginary and the foliation will have to end on that largest of the two roots. So, for the foliation to extend all the way to the singularity and still have a universal horizon one needs to require that
\begin{equation}
	\frac{(4a^2 +\mathcal{J}^2)(b^2-\bL(b))}{\xi(\sM -2ab)^2} = 1.
	\label{eqn:constraintUH}
\end{equation}
We can use this constraint to express $a$ in terms of the other parameters $\{\sM, \mathcal{J},b\}$, thus reducing the dimension of the parameter space to three.

Assuming that the resulting $r_u^2$ is real (which imposes a further constraint on the parameters), the universal horizon is uniquely located at
\begin{equation}
r_u^2= \frac{\mathcal{M}-2a_\pm(\sM, \mathcal{J},b)b}{2(b^2-\bar{\Lambda(b)})}, 
\label{eqn:locUH}
\end{equation}
where $a$ is now understood to depend on the other parameters through Eq. (\ref{eqn:constraintUH}). Because Eq. (\ref{eqn:constraintUH}) is quadratic in $a$, there will generally be two values of $a$ (which we denote by $a_{\pm}$) for every choice of $\{\sM, \mathcal{J},b\}$. Each particular triple $\{\sM, \mathcal{J},b\}$ can represent two distinct solutions, each possibly harboring a universal horizon. 

\subsubsection{Black holes with universal horizons}

\begin{figure}[h]
\begin{center}
 \includegraphics[width=7.5cm]{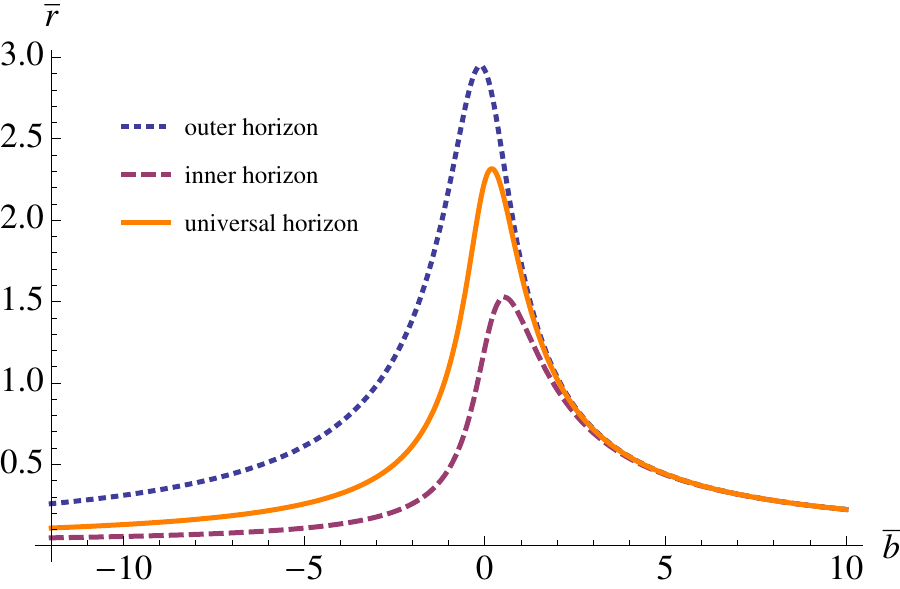}
 \caption{(Color online) Radial positions of various horizons in a BTZ-like anti-de Sitter black hole.}
\label{fig:horizonsAdS}
\end{center}
\end{figure}

For a BTZ-like solution with AdS asymptotics, the universal horizon is located between the outer and inner event horizons. This is illustrated in Fig.~\ref{fig:horizonsAdS}. Note that $b$ and $\mathcal{J}$ are dimensionful quantities ($[b]= 1/L, [\mathcal{J}]=L$); for the plots we use their dimensionless versions $\bar{r} := r/l$ and $\bar{b} := bl$, where $l$ is the ``bare'' cosmological length scale, $l := 1/\sqrt{|\Lambda|}$.  Fig.~\ref{fig:horizonsAdS} shows the positions of the horizons as a function of $\bar{b}$, keeping other parameters fixed at $\{\sM = 10, \mathcal{J}/l=0.1\}$ and with the coupling constants set to be $\{\xi =1/2, \lambda = 1\}$.

\begin{figure}[h]
\begin{center}
 \includegraphics[width=3.8cm]{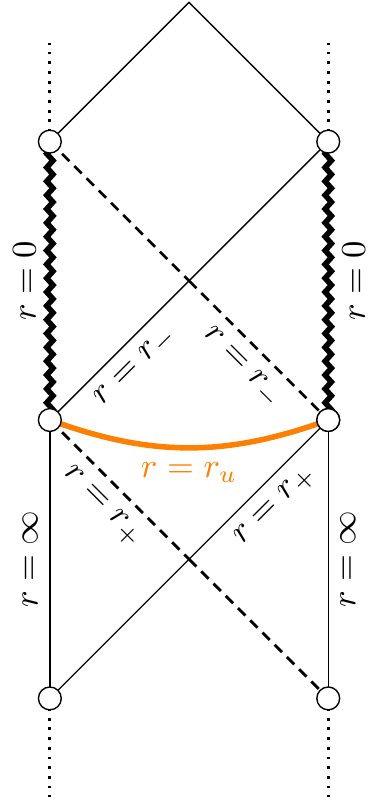}
 \caption{(Color online) Penrose diagram for $\sM >0, \bL <0, \bJ^2>0$. This is equivalent to the rotating BTZ case, except for the curvature singularity at $r=0$. The dashed lines represent null leaves of the foliation. Along these surfaces, the aether diverges because it becomes lightlike. The orange solid curve represents the universal horizon (when it exists).
 }
\label{fig:uhPenroseADS}
\end{center}
\end{figure}

We have also chosen the sign of the bare cosmological constant to be negative, so that $\bL(b=0)< 0$. To ensure that the aether represents a well-defined folation at large $r$ for any value of $b$, we need to work within the parameter region $\{\xi > 0,\lambda > 1/2\}$. With $\Lambda < 0$, any choice from this region guarantees that $(b^2 - \bL(b)) = -\Lambda + b^2(2\lambda-1)/\xi > 0$ is non-negative for any value of $b$. Moreover, if one chooses them such that $\lambda \geq (1+\xi)/2$, then $\bL$ is always negative for any $b$.  Fig.~\ref{fig:horizonsAdS} is such a case, where all values of $b$ give regular AdS black holes. Fig.~\ref{fig:uhPenroseADS} shows the locations of the universal horizon in the Penrose diagram of an AdS black hole spacetime.

\begin{figure}[h]
\begin{center}
 \includegraphics[width=7.5cm]{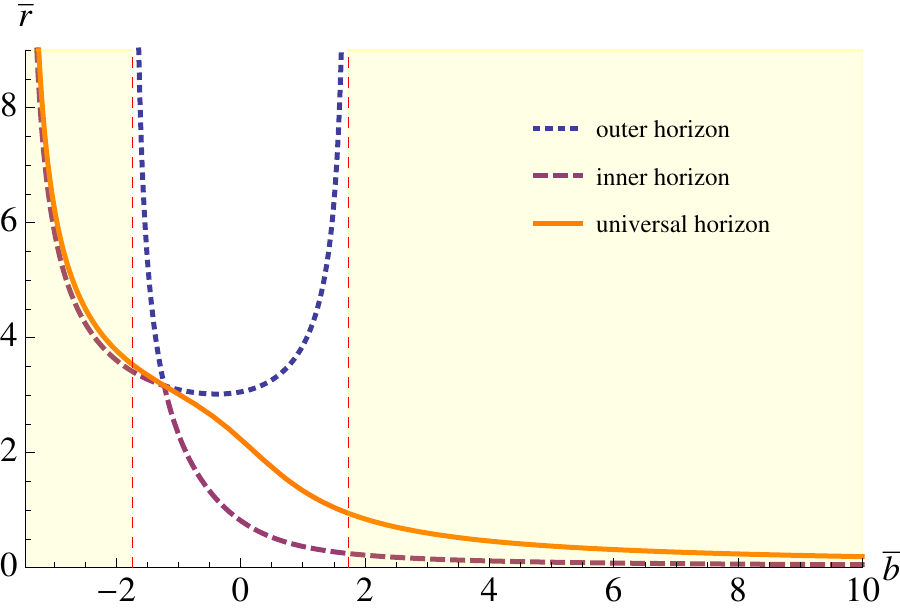}
 \caption{(Color online) Transitioning from AdS to dS asymptotics. The yellow shaded regions are asymptotically de Sitter spacetimes, while the unshaded region represents an AdS black hole.
 }
\label{fig:horizonsAdSDS}
\end{center}
\end{figure}

Now if the coupling constants are such that $\{\xi > 0,\lambda > 1/2\}$ and $\lambda <  (1+\xi)/2$, then $\bL$ will switch sign at some value of $b$. When this happens, the aether charge $b$ radically changes the causal structure of the spacetime. In Fig.~\ref{fig:horizonsAdSDS}, we have an example of a spacetime starting with AdS asymptotics at $b=0$ and turning asymptotically dS as $b$ is increased. This plot is made with the parameters $\{\sM = 10, \mathcal{J}/l=0.1\}$, but with $\{\xi =3/4, \lambda = 3/4\}$. One can verify that the spacetime turns dS at $\bar{b} = \pm\sqrt{3}$. The shaded regions denote solutions that are asymptotically dS, but these solutions are not black holes since $\sM >0$ and $\bJ^2 >0$. (For $0<\xi<1$, $\bJ^2$ is always positive). Only the unshaded regions -- those with AdS asymptotics -- are black holes. 

Interestingly, within the AdS region, there is a kink in the curves, $\bar{r}_\pm(\bar{b})$. For this case, this occurs around $\bar{b} = -1.2247$, which is where $1+\bJ^2 \bL/\sM^2$ vanishes. We note that while both curves touch, they do not cross over. At this point, which is also where all horizons meet, $\bar{r}_\pm(\bar{b})$ are continuous but not differentiable with respect to the parameter $\bar{b}$.

As the transition from AdS to dS asymptotics is made, the outer horizon is pushed to $r=\infty$, leaving as the ``outer'' region of the asymptotically dS spacetime what was formerly the interior of the AdS black hole. At the same time, the inner horizon of the AdS black hole turns into the dS event horizon. The universal horizon remains in between the inner and outer horizons of the AdS black hole, and can be found in the ``outer'' region of the dS spacetime. 

That the universal horizon tends to be located beyond the dS horizon (i.e., at a larger value of $r$) appears to be a generic property of these solutions. Such a horizon can be thought of as a {\em cosmological universal horizon}.

It is also of interest to look at the case of the dS black hole. Choosing the sign of the bare cosmological constant to be positive this time ($\Lambda >0$), we now choose the other parameters to be $\{\sM = -10, \mathcal{J}=0.1\}$ and the coupling constants $\{\xi =2, \lambda = 1\}$. The coupling constants are chosen so that all values of $b$ lead to dS asymptotics, which is $\lambda < (1+\xi)/2$ for $\Lambda >0$. However, to guarantee that the aether is real at large $r$ [re: $b^2 > \bL(b)$], we are limited to the region $\bar{b} \geq \sqrt{2}$. 

\begin{figure}[h]
\begin{center}
 \includegraphics[width=7.5cm]{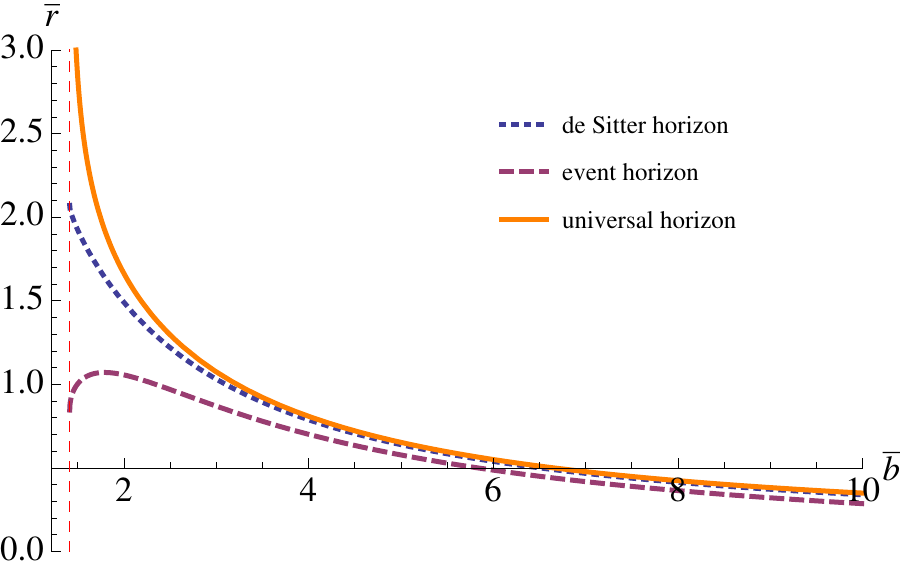}
 \caption{(Color online) Radial positions of various horizons in a de Sitter black hole.}
\label{fig:horizonsDS}
\end{center}
\end{figure}

\begin{figure}[h]
\begin{center}
 \includegraphics[width=7.5cm]{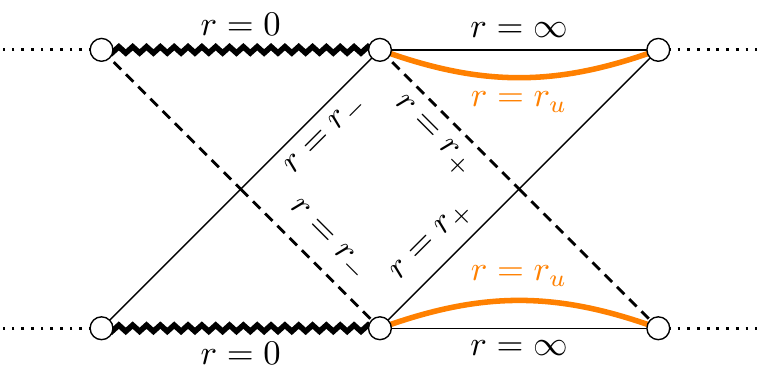}
 \caption{(Color online) Penrose diagram for $\sM <0, \bL > 0, \bJ^2 <0$. This is equivalent to the Penrose diagram for Schwarzschild-de Sitter spacetime. The dashed lines represent null leaves of the foliation. The orange solid curves represent universal horizons (when they exist).}
\label{fig:uhPenroseDS}
\end{center}
\end{figure}

For all values of $\bar{b}$ shown in Fig.~\ref{fig:horizonsDS}, the spacetime is a dS black hole with an event horizon and a dS horizon. However, for sufficiently large $\bar{b}$ (not shown in the plot), $\bJ^2$ becomes positive, and the event horizon ceases to exist. Again, we see here that the universal horizon is located beyond the dS horizon. In Fig.~\ref{fig:uhPenroseDS}, the universal horizon is displayed in the Penrose diagram of a dS black hole spacetime.

\begin{figure}[h]
\begin{center}
 \includegraphics[width=6cm]{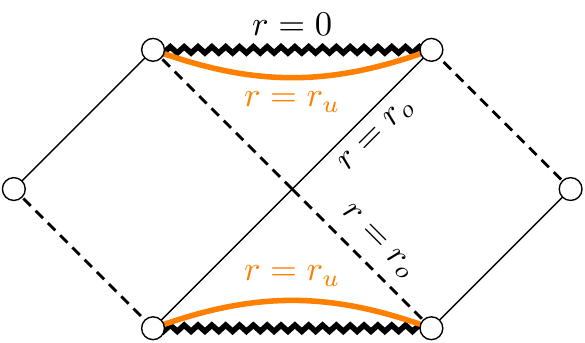}
 \caption{(Color online) Penrose diagram for an asymptotically flat black hole, whose causal structure is essentially that of the Schwarzschild spacetime. The dashed line represents a null leaf of the foliation. The orange solid curves represent universal horizons (when they exist).}
\label{fig:uhPenroseFlat}
\end{center}
\end{figure}

\begin{figure}[h]
\begin{center}
 \includegraphics[width=3.8cm]{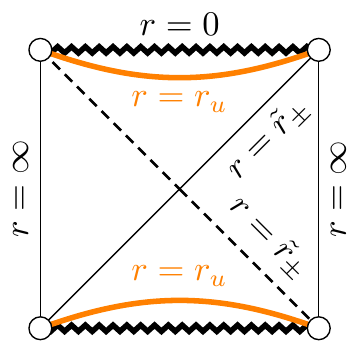}
 \caption{(Color online) Penrose diagram for $\bL <0, \bJ^2<0$, and for either sign of $\sM$ ($r=\tilde{r}_+$ for $\sM >0$ and $r = \tilde{r}_-$ for $\sM < 0$). The dashed line represents a null leaf of the foliation. The orange solid curves represent universal horizons (when they exist).}
\label{fig:PenroseOneKillingADS}
\end{center}
\end{figure}

\begin{figure}[h]
\begin{center}
 \includegraphics[width=3.9cm]{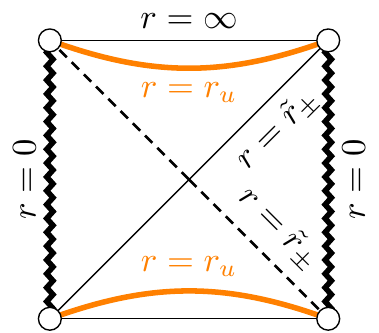}
 \caption{(Color online) Penrose diagram for $\bL > 0, \bJ^2 > 0$ and for either sign of $\sM$  ($r=\tilde{r}_+$ for $\sM < 0$ and $r = \tilde{r}_-$ for $\sM > 0$). The dashed line represents a null leaf of the foliation. The orange solid curves represent universal horizons (when they exist).}
\label{fig:PenroseOneKillingDS}
\end{center}
\end{figure}

In the asymptotically flat case $\bL = 0$, the aether charge $b$ is fixed at a particular value: 
\begin{equation}
\label{blambda}
	b_{\rm flat}^2 = \Lambda \left(\frac{\xi}{2\lambda -\xi -1}\right). 
\end{equation} 
It is quite straightforward to choose parameters for which the universal horizon exists. Asymptotically flat solutions with universal horizons have no extra hair (i.e. independent aether charge) apart from $\sM$ and $\mathcal{J}$.
In Fig. \ref{fig:uhPenroseFlat}, the universal horizon is displayed in the Penrose diagram of an asymptotically flat, black hole spacetime. Finally, Figs. \ref{fig:PenroseOneKillingADS} and \ref{fig:PenroseOneKillingDS} are Penrose diagrams for spacetimes with only one Killing horizon, the former being a spacetime with a black hole horizon and the latter a cosmological horizon. 

\subsubsection{Black holes without universal horizons}

We have implicitly already stated two conditions for universal horizons to not exist at all: firstly, the discriminant in Eq. (\ref{eqn:rootUH}) can be negative, and secondly, $r_u^2$ can be negative. It is worth pointing out that these condition can be satisfied even in black hole solutions if the parameters are chosen appropriately.

Consider, as an example, the black hole with flat asymptotics, $\{\bL =0, \sM =-1, \bJ^2 < 0\}$, and assume, additionally, that $b=0$ so that the aether asymptotically aligns with the timelike Killing vector. Eq.~(\ref{blambda}) requires that $\Lambda$ has to vanish as well. One can then straightforwardly calculate the root of Eq.~(\ref{eqn:univhor}). This is
\begin{equation}
r_u^2=-\left(\frac{J^2+4a^2}{\xi}\right),
\end{equation}
and it is negative-definite ($\bar{J}^2<0$ requires that $\xi>1$). So, no universal horizon exists for black holes with flat asymptotics \emph{and} an aether that asymptotically aligns with the timelike Killing vector.

As another example, let us consider black holes with AdS asymptotics. The negative discriminant condition reads
\begin{equation}
	\frac{(4a^2 +\mathcal{J}^2)(b^2-\bL(b))}{\xi(\sM -2ab)^2} > 1\,,
\label{eqn:noUH}
\end{equation}
while the black hole bound given in Eq. (\ref{eqn:bound}) for $\bL(b) < 0$ is
\begin{equation}
0 \leq \mathcal{M}^2 + \bar{\Lambda}(b)\bar{\mathcal{J}}(a)^2.
\label{eqn:bound2}
\end{equation}
Finally, we also need to require that the aether is real at large $r$ ($b^2 \geq \bL(b)$) and small $r$ ($\xi>0$).
All need to be satisfied for the parameters to represent regular black hole solutions without universal horizons.

\begin{figure}[h]
\begin{center}
 \includegraphics[width=7.5cm]{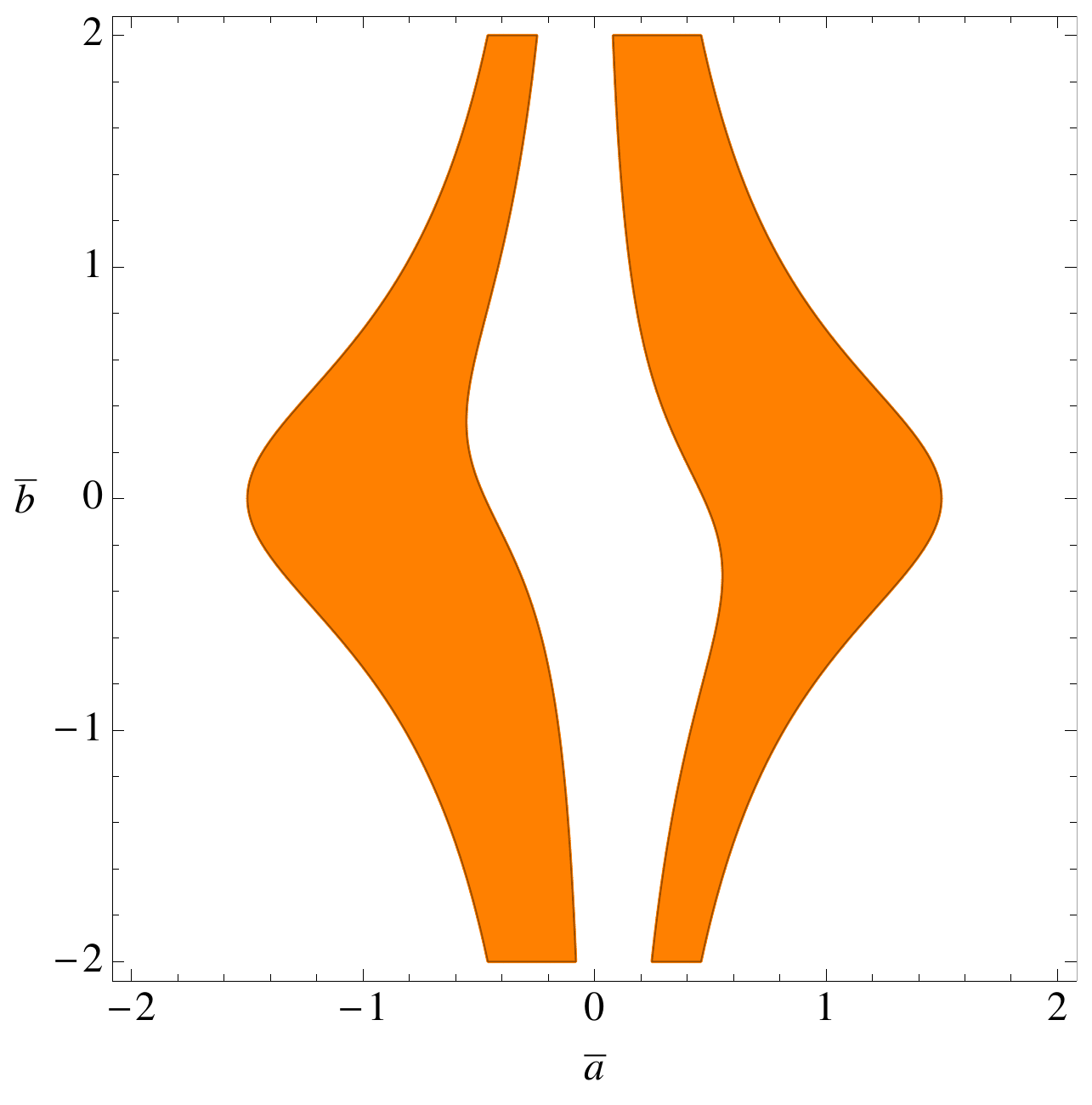}
 \caption{(Color online) Sample region of parameter space representing to asymptotically anti-de Sitter black holes without universal horizons. These plots use $\{\sM = 1,\Lambda l^2 = -1, \mathcal{J}/l=1\}$, $\lambda =2$, and $\xi = 0.9$. Here, $\bar{a} := a/l$ and $\bar{b} := bl$, with $l:= 1/\sqrt{-\Lambda}$.}
 \label{fig:UHnew}
\end{center}
\end{figure}

We graphically demonstrate that a fairly large region of parameter space satisfies all these requirements. For the values $\{\sM = 1,\Lambda l^2 = -1, \mathcal{J}/l=1\}$, $\lambda =2$ and $\xi = 0.9$, we display in Fig. \ref{fig:UHnew} the values of $\{a,b\}$ satisfying (a) $\bL<0$, (b) the black hole bound in Eq. (\ref{eqn:bound}), (c) the negative discriminant condition in Eq. (\ref{eqn:noUH}), and (d) the aether regularity constraint at large $r$. These all correspond to asymptotically AdS black holes with no universal horizons. 

\subsection{Nonrotating limit}

One can choose to focus in the ${\mathcal J}=0$ case which corresponds to a nonrotating black hole. In general the spacetime retains most of the features it had when ${\mathcal J}\neq 0$ provided that $a\neq 0$. Curvature invariants still diverge at $r=0$ and the causal structure remains largely unaffected. Ergosurfaces now coincide with the metric horizons, as expected.
Nevertheless, it is worth pointing out that one can still have two black hole horizons in black hole solutions with AdS asymptotics.

As far as  universal horizons are concerned, they can be present or absent, depending on the solutions. 
When $\mathcal{J} = 0$, the constraint given by Eq. (\ref{eqn:constraintUH}) reduces to
\begin{equation}
	\frac{4a^2(b^2-\bL(b))}{\xi(\sM -2ab)^2} = 1.
\end{equation}
One can readily identify two characteristic examples of nonrotating black holes that cannot satisfy this constraint and cannot have a universal horizon. The first is the asymptotically flat black hole with $b=0$ (discussed above) and $J=0$. The second is a black hole with AdS asymptotics and $a=0$. This is actually a nonrotating BTZ black hole with a nontrivial aether configuration.

\section{Conclusions}

Our intention was to find an analogue of the BTZ black hole in three-dimensional Ho\v rava gravity. To this end we first considered whether AdS space or AdS asymptotics are admissible in this theory. Using the reduced action approach we have shown that this is only true if $\eta=0$. We subsequently focused on the $\eta=0$ sector of the theory. We have found the most general class of solutions in this sector, without imposing specific asymptotics.  Remarkably, the black hole solutions in this class do not have exclusively AdS asymptotics, but there exist instead also black holes with dS and flat asymptotics, unlike general relativity.

The black hole solutions we found have very interesting properties. They harbor a curvature singularity, unlike their GR counterparts. They can have an inner and an outer metric (Killing) horizon and one or two ergosurfaces. What is perhaps their most interesting feature within the context of Lorentz-violating gravity theories is that they can have universal horizons. Rotation does not seem to play a key role in the existence of these horizons. Depending on the configuration of the preferred foliation, there can be nonrotating black holes without universal horizons or rapidly rotating black holes with universal horizons. Some of our solutions also feature the existence of cosmological universal horizons. These results demonstrate that the existence of universal horizons does not seem to depend on spherical symmetry or the number of spacetime dimensions and it is not specific to black hole spacetimes. At the same time, they also highlight the importance of the asymptotic behavior of the foliation for the existence of universal horizons.

The $\eta=0$ sector of three-dimensional Ho\v rava gravity, to which the requirement of AdS asymptotics has restricted us, is likely to be a special theory. At the perturbative level the scalar mode that the theory propagates appears to travel at infinite speed and, at the same time, the theory is expected to be strongly coupled \cite{Sotiriou:2011dr}. In four dimensions choosing $\eta=0$ leads to a physically (but not mathematically) inconsistent theory \cite{Henneaux:2009zb}. Nevertheless, we expect the black hole solutions we present here to be useful tools for applications such as quantum field theory near horizons in the presence of Lorentz violations and black hole thermodynamics, so long as one remains cautious about the interpretation of the results.

Finally, the existence of black hole solutions with flat or dS asymptotics in the $\eta=0$ sector of the theory suggests that it is also likely for black hole solutions with these asymptotics to exist when $\eta\neq 0$. We shall explore this possibility in future work. 

\begin{acknowledgments}
The authors thank Jishnu Bhattacharyya and Ted Jacobson for fruitful discussions that helped shape this work. The research leading to these results has received funding from the European Research Council under the European Union's Seventh Framework Programme (FP7/2007-2013)/ERC Grant Agreement n. 306425 ``Challenging General Relativity''.
\end{acknowledgments}

\appendix

\section{Energy conditions}
\label{app:energy}

It is convenient to check energy conditions using the preferred rest frame of the aether. The weak energy condition (WEC) states that the energy density measured by an arbitrary observer must be positive. Taking this observer to be at rest with respect to the aether, we have
\begin{equation}
	T^{\alpha\beta}u_\alpha u_\beta = \frac{\bJ^2-\mathcal{J}^2+ 4\bL r^4}{4r^4}\geq 0, 
\end{equation} 
where $T^{\alpha\beta} := R^{\alpha\beta}-(1/2)g^{\alpha\beta}R$, 
and $u_{\alpha}$ is the aether vector field.
 
Insisting that this condition is satisfied for all $r$ requires $\bL>0$ and $\bJ^2 \geq \mathcal{J}^2$. AdS asymptotics thus violates the WEC. On the other hand, in order to have black hole horizons in solutions with dS or flat asymptotics, $\bJ^2$ has to be negative. This means that the WEC is violated in these solutions as well. Hence, all our black hole solutions violate the WEC. Since the WEC is a necessary condition for the dominant energy condition (DEC), all our black holes violate the DEC as well.

Violating the DEC is to be expected from the work of Ida \cite{Ida:2000jh}, which states that if the DEC is satisfied then the spacetime cannot have apparent horizons. Since apparent horizons are also event horizons in stationary spacetimes, this result precludes the existence of black holes when the DEC holds. 

\section{Aether alignment} 
\label{app:align}

Because the timelike Killing vector, $t^\alpha := (\frac{\partial}{\partial t})^\alpha$, turns null in black hole spacetimes, the aether cannot be aligned with it everywhere. In this appendix, we work out how alignment is realized in terms of our unknown functions, $\{Z,F,U, \Omega\}$. If $u^\alpha$ is aligned with $t^\alpha$, then $t_\alpha a^\alpha = 0$, where $a^\alpha := u^\beta\nabla_\beta u^\alpha$. Normalizing $t^\alpha$ to get $\hat{t}^\alpha := t^\alpha/\sqrt{g_{\mu\nu}t^\mu t^\nu}$, alignment is then equivalent to 
\begin{equation}
	\hat{t}^\alpha a_\alpha = - \frac{U F^3 \left((U^2F^2+1)Z^2\right)'}{Z \sqrt{F^2(1 + U^2F^2)}\sqrt{Z^2 -r^2 \Omega^2}}=0.
	\label{eqn:alignCond}
\end{equation}
This is satisfied either when $F=0$, $U=0$, or 
\begin{equation}
	\left(U^2F^2 + 1\right)Z^2 = C^2 \,\,\,\Longrightarrow\,\,\, u_t = C,
\end{equation}
for some constant $C$.
This latter case just corresponds to a zero-acceleration aether, where the foliation is provided by the Painleve-Gullstrand time. It does not represent alignment between the aether and the timelike Killing vector. Therefore, since $F$ cannot vanish everywhere, the timelike Killing vector and the aether are aligned everywhere if and only if $U=0$.

Now specializing to our solution, where $Z=F$, we can use Eq. (\ref{eqn:alignCond}) to check for asymptotic alignment. For AdS asymptotics, $F^2 \sim r^2$. As we have shown in Sec. \ref{sec:maxSymm} and Appendix \ref{app:asympAdS}, the aether component $u_r$ (or $U$) can then only fall-off as $U \sim r^{-1}$ or $U \sim r^{-3}$. It is easy to check that $\hat{t}^\alpha a_\alpha \sim r^0$ or $\hat{t}^\alpha a_\alpha \sim r^{-2}$, respectively. It aligns asymptotically only for the latter case. Thus, when $b \neq 0$ in Eq. (\ref{eqn:fol}), and hence $U \sim r^{-1}$, the aether does not become orthogonal to constant-$t$ surfaces as $r\rightarrow \infty$. The parameter $b$ is then a measure of the asymptotic misalignment of the aether.

\section{Metric ansatz in the preferred time}
\label{app:aligned}

The approach in \cite{Park:2012ev} amounts to setting $\eta=0$ and $U=0$ in Eq. (\ref{eqn:reducedL})\footnote{Reference \cite{Park:2012ev} also includes an $R^2$ term in the action. We do not include this term and work fully within the infrared sector.}. Doing so gives the much simplified Lagrangian
\begin{equation}
    L_{\rm align} =  \frac{r^3F}{2Z}({\Omega}'){}^2 - 2\xi Z\left(\Lambda\frac{r}{F}+F'\right).
\end{equation}
However, the metric ansatz given in (\ref{eqn:btzansatz}) is not the most general stationary metric when one works within the preferred foliation.

By fiat, the aether is normal to constant-$T$ surfaces and thus $u_{T}$ will be the only nonvanishing component in the preferred foliation. To bring the metric ansatz of Eq. (\ref{eqn:btzansatz}) into the preferred frame, one needs to perform the coordinate transformation that puts the aether into this form. 
 Explicitly, this is 
\begin{eqnarray}
T &= t + \int^r \frac{u_r(r')}{u_t(r')} dr'\,.
\end{eqnarray}
Thus, the metric ansatz in the preferred frame in terms of the unknown functions $Z,F,\Omega$ and $U$ is
\begin{align}
ds^2 = Z^2 &dT^2 -\frac{2ZU}{\sqrt{1+F^2U^2}}dTdr -\frac{1}{F^2(1+F^2U^2)}dr^2 \nonumber \\ &-r^2\left(d\phi+\Omega dT - \frac{\Omega U}{Z\sqrt{1+F^2U^2}}dr \right)^2.
\end{align} 
Inserting this metric ansatz directly into the preferred frame action Eq. (\ref{eqn:hlaction}) provides an equivalent strategy to the one we have adopted.

The metric in the preferred foliation will generally have a $g_{Tr}$ and a $g_{r\phi}$ component because the aether will not be orthogonal to constant-$t$ hypersurfaces, or equivalently, $T$ and $t$ do not generally coincide. Evidently, an aligned aether configuration is just a special case, which in our parametrization is $U=0$.

\section{Constant-$T$ surfaces have constant mean curvature when $\eta=0$}
\label{app:meancurv}

In this appendix we derive the mean curvature, $K$, of a constant-$T$ surface. The extrinsic curvature is defined as 
\begin{equation}
K_{\alpha\beta} := h_\alpha{}^\gamma h_\beta{}^\delta \nabla_\gamma u_\delta,
\end{equation}
where the $h_\alpha{}^\beta := g^{\beta\delta} h_{\alpha\delta} = g^{\beta\delta} (g_{\alpha\delta}-u_\alpha u_\delta)$ are spatial projectors. The mean curvature is then just $K:= g^{\alpha\beta}K_{\alpha\beta}$.

In terms of the functions $\{Z,F,U\}$, the mean curvature can be written as
\begin{equation}
K = -U F^2 \left(\frac{d}{dr}(\log ZFU)+ \frac{1}{r}\right).
\label{eqn:meangeneral}
\end{equation} 
A straightforward calculation then reveals that the aether field in Eq. (\ref{eqn:fol}) defines a surface with constant mean curvature $K=-2b$. This turns out to be a necessary condition for any $\eta = 0$ solution.

Because $\eta = 0$ implies $F=Z$, in our parametrization the mean curvature according to Eq. (\ref{eqn:meangeneral}) is just
\begin{align}
K= -UF^2\left(\frac{d}{dr}(\log UF^2) + \frac{1}{r}\right) &= -y \left(\frac{d}{dr}\log y + \frac{1}{r}\right) \nonumber \\ &= -\Bigg(y' + \frac{y}{r}\Bigg),
\end{align} 
where again we have used the substitution in Eq. (\ref{eqn:subs}), $y=UF^2$. Therefore, 
\begin{equation}
r^2 K' = -(r^2y''+ry'-y),
\end{equation}
so Eq. (\ref{eqn:simple}) is equivalent to $K'=0$ or $K=$ constant. In other words, when $\eta=0$, the aether defines surfaces of constant mean curvature.

\section{Brown-Henneaux asymptotic conditions for anti-de-Sitter spacetime}
\label{app:asympAdS}

Inserting Brown-Henneaux boundary conditions into the EL equations results in rather complicated expressions, but our interest here is to investigate only the leading-order terms. For this it will suffice to consider just the numerators of the expressions. For example, consider the expression
\begin{equation}
	f:= \frac{ar^k + br^{(k-1)}+\cdots}{cr^j + dr^{(j-1)}+\cdots},
\end{equation}
with $k>0$ and $j>0$ (for the sake of argument). Then as $r\rightarrow \infty$, the leading-order term of $f$ is 
\begin{equation}
	\frac{ar^k + br^{(k-1)}+\cdots}{cr^j + dr^{(j-1)}+\cdots} \sim \frac{a}{c}r^{(k-j)}.
\end{equation}
Enforcing that $f$ vanishes asymptotically to leading order requires only that $a=0$, so it is sufficient to focus mainly on the numerator of $f$. We shall call this the leading-order coefficient (LOC). The denominator merely rescales the LOC by a constant and so it shall not play an important role in the leading-order asymptotic analysis.

The LOCs of the asymptotic EL equations will depend on $m$. For the $Z$ equation, the dominant term in the numerator is either $\sim r^{12}$ or $\sim r^{(4m+16)}$. When $m< -1$, $r^{12}$ dominates. When $m > -1$, $r^{(4m+16)}$ dominates. And when $m=-1$, both terms (along with several others) scale with $r$ in the same way (i.e., $\sim r^{12}$).

For the $F$ equation, you get something similar. The dominant term in the numerator is either $\sim r^{14}$ or $\sim r^{(6m+20)}$. When $m< -1$, $r^{14}$ dominates. When $m > -1$, $r^{(6m+20)}$ dominates. And when $m=-1$, both terms (along with several others) scale with $r$ in the same way (i.e., $\sim r^{14}$).

Finally, for the $U$ equation, the dominant term in the numerator is either $\sim r^{(m+14)}$ or $\sim r^{(5m+18)}$. When $m< -1$, $r^{(m+14)}$ dominates. When $m > -1$, $r^{(5m+18)}$ dominates. And when $m=-1$, both terms (along with several others) scale with $r$ in the same way (i.e., $\sim r^{13}$).

Clearly, $m=-1$ is the critical value for the analysis. We shall investigate each of the cases in turn: $\{m > -1, m=-1, m<-1\}$

\subsection{Case I: $m > -1$}

In this case, the LOCs of the $U$, $F$, and $Z$ equations (modulo harmless factors) are, respectively, 
\begin{multline}
	(3+4\eta -3\lambda) + 4(1+\eta-\lambda)m + (1+\eta-\lambda)m^2, \\
	(1+4\eta +3\lambda) + \left[4\eta-2(-2+\lambda+\xi)\right]m + (1+\eta-\lambda)m^2, \\
	\Big(\frac{11}{3}+4\eta -5\lambda + \frac{4}{3}\xi\Big) + \frac{2}{3}(6+6\eta-7\lambda+\xi)m + (1+\eta-\lambda)m^2.
\end{multline}
These clearly do not vanish simultaneously for generic coupling constants. For them to vanish simultaneously, the coupling constants will have to be especially chosen. This can only happen if the expressions are identical. The coefficients of the terms linear in $m$ have to match. So, 
\begin{equation}
4(1+\eta-\lambda) = \left[4\eta-2(-2+\lambda+\xi)\right] = \frac{2}{3}(6+6\eta-7\lambda+\xi).
\end{equation}
This is a system of three equations in three unknowns, for which the solution is simply 
\begin{equation}
	\xi = \lambda. \label{eqn:sol1}
\end{equation}
The constant terms [i.e. $O(m^0)$] also have to match
\begin{equation}
(3+4\eta -3\lambda) = (1+4\eta +3\lambda) = \left(\frac{11}{3}+4\eta -5\lambda + \frac{4}{3}\xi\right),
\end{equation}
which gives
\begin{align}
	\xi &= 0 \\
	\lambda &= \frac{1}{3}. 
\end{align}
Because these are incompatible with Eq. (\ref{eqn:sol1}), we conclude that the leading-order terms of the EL equations cannot simultaneously vanish. This means that for $m > -1$, in $U \sim U_0r^m$, AdS asymptotics for the metric are inadmissible.

\subsection{Case II: $m = -1$}

In this case, enforcing that the LOCs of the $U$, $F$, and $Z$ equations vanish (respectively), we have 
\begin{multline}
	\eta U_0 \Big(\sL^2 + U_0^2\Big)^2 = 0, \\ 
	2 \sL^4 \xi + (1/\Lambda) (\sL^2 (\eta + 2 \xi) - (-2 + \eta + 4 \lambda - 2 \xi) U_0^2) = 0, \\
	2 \sL^4 \xi + (1/\Lambda) (\sL^2 (3 \eta + 2 \xi) + (2 + 3 \eta - 4 \lambda - 2 \xi) U_0^2) = 0. 
\end{multline}

The first of these equations demands that $\eta =0$. When this is the case, the other two equations lead to the same solution
\begin{equation}
U_0^2 = \sL^2\left(1 + \Lambda \sL^2 \right)\frac{\xi}{2\lambda - \xi-1}. 
\label{eqn:sol2}
\end{equation}
This we can verify to be the first aether charge of our solution. As $r \rightarrow \infty$, our exact solution behaves like
\begin{equation}
	U \sim \frac{br}{F^2} \sim \frac{br}{r^2/\sL^2} = \frac{b\sL^2}{r},
\end{equation}
taking note of the fact that $\bL = -1/\sL^2$.
Therefore, since in our asymptotic analysis, $U\sim U_0/r$ (for $m=-1$), we must have $U_0 = b\sL^2$. 

On the other hand, from Eq. (\ref{eqn:effLambda}) we have
\begin{align}
		  \bL &= \Lambda - \frac{b^2(2\lambda-\xi-1)}{\xi},\\
		  		  -\frac{1}{\sL^2} &= \Lambda  - \frac{U_0^2}{\sL^4}\frac{(2\lambda-\xi-1)}{\xi},
\end{align}
which is identical to Eq. (\ref{eqn:sol2}). This demonstrates that the asymptotic analysis recovers one of the aether charges (i.e. $b$) for the case $m=-1$. What is most essential, however, is that $m=-1$ forces us to set $\eta =0$.

\subsection{Case III: $m < -1$}

For this final case, the LOCs of the $U$, $F$ and $Z$ equations give
\begin{align}
	(3 + \eta - 3\lambda) + 4(1 - \lambda)m + (1 - \lambda)m^2 &= 0, \nonumber \\
	2 \sL^2 \xi + (1/\Lambda) (\eta + 2 \xi) &= 0, \nonumber \\
	2 \sL^2 \xi + (1/\Lambda) (3\eta + 2 \xi) &= 0. 
\end{align}
The second and third of these equations imply again that $\eta$ has to be zero. Putting this into the first equation gives 
\begin{equation}
	(m+1)(m+3)(-1 + \lambda) = 0.
\end{equation}
Since $m<-1$ and since we wish to keep the coupling constants as generic as possible, we must choose $m=-3$. Moreover, the second and third equations give
\begin{equation}
	2\xi(\sL^2+1/\Lambda) = 0 \Longrightarrow \sL^2 = -1/\Lambda.
\end{equation}
In other words, the effective cosmological constant must be the bare one: $\bL = \Lambda$. Again, however, this case shows that $\eta = 0$ is required.

To summarize, we have demonstrated in this appendix that Brown-Henneaux AdS boundary conditions forces us into the $\eta=0$ sector. As an added bonus, we see that for AdS asymptotics, $U$ can only scale as $r^{-1}$ or $r^{-3}$ at large values of $r$, indicating the existence of two asymptotic aether charges, which is precisely what we find in our exact solution.

\section{Special choices of Horava parameters}
\label{app:lambdaone}

Within the $\eta=0$ sector, $\lambda =1$ is  special because we lose the constraint provided by Eq. (\ref{eqn:ueqn}). The $U$ equation is identically satisfied and one is left with an underdetermined system for the functions $U$ and $F$.

The $Z$ and $F$ equations provide the sole constraint:
\begin{align}
(\xi -1)\Big[&\frac{d}{dr}\left(U^2\right) + 4\left(\frac{F'}{F}\right)U^2 \Big] \nonumber \\ &+ 2\xi\frac{F'}{F^3} + \left(\frac{\mathcal{J}^2}{2r^3}+2r\xi\Lambda\right)\frac{1}{F^4} = 0 
\end{align}
which can be integrated to give
\begin{equation}
(\xi-1)U^2 = \frac{1}{F^2}\left[\mathcal{C}+\frac{\mathcal{J}^2}{4 r^2} - \xi\left(\Lambda r^2 + F^2\right)\right]
 \label{eqn:lambdaone}
\end{equation}
for some integration constant $\mathcal{C}$. When $\xi =1$,  Eq. (\ref{eqn:lambdaone}) does not depend on $U$ and becomes purely a condition on $F$. In this case, it returns for $F$ the BTZ solution of general relativity, while $U$ can be any function. This result is not surprising. For $\eta=0$, $\lambda=\xi=1$ Ho\v rava gravity in its covariant version is equivalent to general relativity with a hypersurface-orthogonal aether that only needs to satisfy the unit constraint without further dynamical restrictions. With our definitions the aether is indeed unity for an arbitrary $U$.

When $\xi\neq 1$, and
since there are no more equations to satisfy, the functions $F$ and $U$ can be chosen so long as they are related according to Eq. (\ref{eqn:lambdaone}). One can verify that no extra conditions arise when working with the full set of field equations instead of the reduced action equations of motion. Note that the condition between $U$ and $F$ is different from Eq. (\ref{eqn:fol}). This result is consistent with the discussion in Sec. \ref{xi=1} about metric and aether redefinitions that set $\xi=1$. One could think of generating the solution for an arbitrary $\xi$ from a solution of the $\xi=1$ theory by an inverse redefinition. Then, a suitable choice of $U$ could lead to the desired $F$.

\end{document}